# Anomalous quantum Hall effect in epitaxial graphene

Xiaosong Wu[1], Yike Hu[1], Ming Ruan[1], Nerasoa K Madiomanana[1], John Hankinson[1], Mike Sprinkle[1], Claire Berger[1,2], Walt A. de Heer[1]

[1] School of Physics / Georgia Institute of Technology, Atlanta, GA-30332, USA
[2] CNRS-Institut Néel, BP 166, 38042 Grenoble Cedex9, France

The remarkable properties of wafer-sized epitaxial graphene (EG) grown on silicon carbide, like its high mobility and graphene electronic structure (*1, 2*) and the fact that it can be patterned, have made it a promising platform for graphene-based electronics (*3*). The observation of the anomalous QHE in microscopic exfoliated graphene flakes that were deposited on silicon oxide substrates (*4*) was pivotal for graphene research. However, the quantum Hall effect (QHE) was elusive in EG, which led to speculations about the quality of EG and the effect of the silicon carbide substrate on transport. The demonstration of the QHE in patterned EG is an important milestone in graphene science. Moreover, its insensitivity to processing induced disorder is significant for the development of epitaxial graphene-based electronics.

Figure 1 shows the magnetotransport measurements of an *EG* monolayer Hall bar sample. The Hall resistance $\rho_{xy}$ clearly exhibits characteristic quantum Hall plateaus that correspond to minima in the longitudinal resisitivity $\rho_{xx}$. The effect is similar to that observed in graphene flakes. Details are given in the figure caption.

A EG monolayer was grown on the C-face of a semi-insulating 4H-silicon carbide substrate (*5,6*) and it was characterized by atomic force microscopy (AFM), ellipsometry and Raman spectroscopy. The EG layer was coated with a PMMA resist and electron-beam patterned as previously described (*1,7*) to produce a typical Hall bar structure as shown in Fig. 1. After patterning, the resist was removed in an (potentially damaging) ultrasonic acetone bath and metal contact pads were applied. The graphene layer charge density corresponds to $n_s=1.2 \times 10^{12}$ electrons/cm$^2$ i.e. a factor of 4 smaller than the charge transfer from the substrate (*5*). This indicates that environmental effects (impurities) significantly reduce the total charge of the EG. Sporadic charging is also seen in exfoliated graphene flakes.

AFM images of the graphene surface show considerable contamination including large PMMA residue particles. The graphene layer has pleats and is draped over several substrate steps. Despite these distortions, the QHE is well developed and the carrier mobility $\mu=19,000$ cm$^2$/Vs (14,300 cm$^2$/Vs at room temperature) is approximately 100 times larger than in equivalently-doped silicon and 10 times larger than that for monolayer epitaxial graphene grown on the Si-face (*3,6*). The role of disorder in the QHE and in graphene transport in general is actively debated (*8*).

We note that distorted QHE-like plateaus have been seen in narrow multilayered epitaxial graphene (MEG) ribbons (*1*), possibly because the undoped graphene overlayers shorted the Hall effect of the conducting interface layer. However, Hall plateaus and magnetoresisance oscillations are all but absent in extended (quasi-two dimensional) MEG samples (*5*). A theoretical explanation has been offered (*9*).

Besides the QHE, we have shown that 1. Epitaxial graphene monolayers can be grown on the C-face of hexagonal silicon carbide wafers; 2. the monolayer is continuous over substrate steps; 3. the mobility is high, despite significant contamination, substrate steps, and harsh processing procedures. 4. The QHE in EG demonstrates that the substrate is at least as unimportant here as it is for "isolated" graphene flakes on silicon dioxide.

Concluding, the robustness and large scale patterning that is possible with epitaxial graphene opens new avenues for graphene physics (*7,10*). This important development brings epitaxial graphene yet a step closer to becoming a scalable platform for graphene-based electronics (*3*).

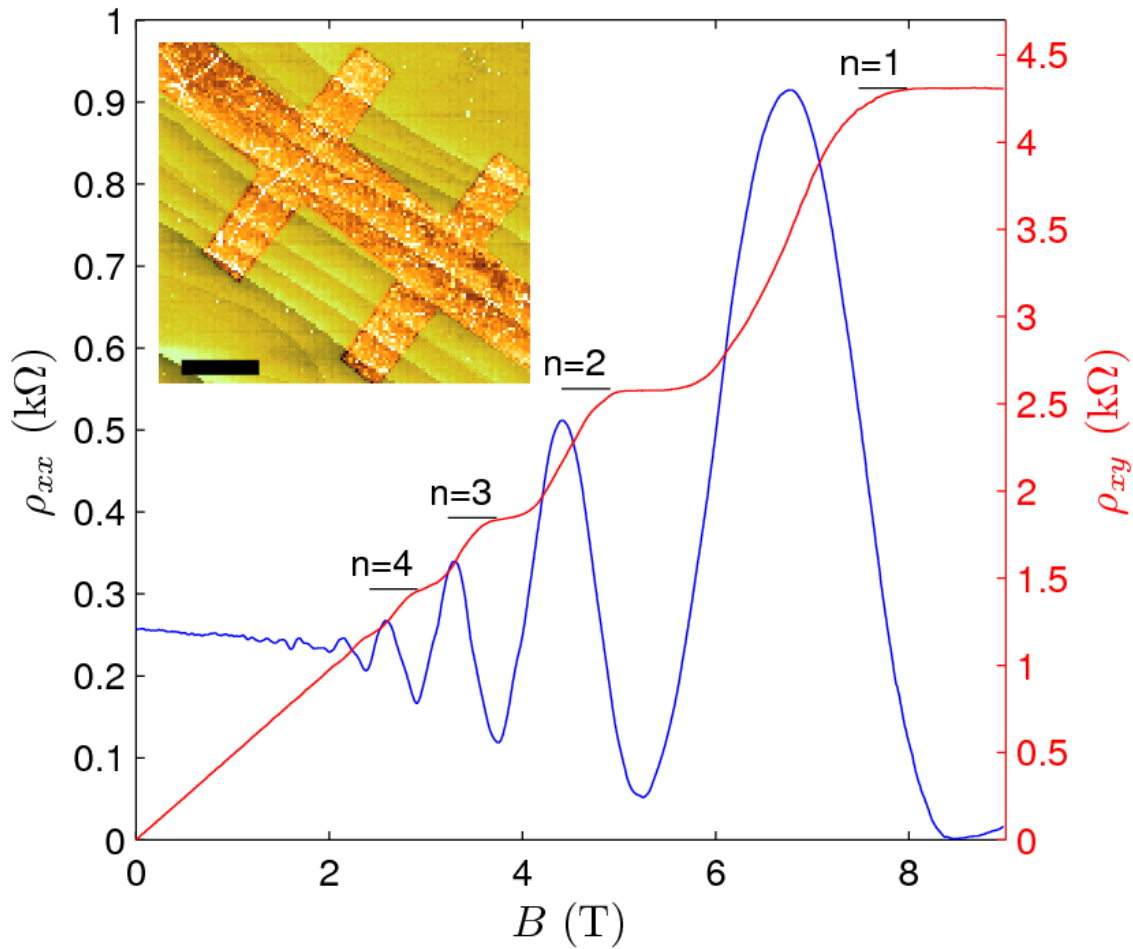

**Figure 1** Quantum Hall effect in monolayer epitaxial graphene measured at 1.4K. Hall resistance (red) as a function of magnetic field showing characteristic Hall plateaus at $\rho_{xy}=(h/4e^2)/(n+1/2)$ where n is the Landau level index ( the n=0 quantum Hall step has been seen in other samples.) , magntoresistance $\rho_{xx}$ (blue) showing characteristic oscillations. The band velocity $v_0=1.14\times10^6$ m/s. Inset : AFM image of the Hall bar (1.8μm x 4.6 μm) patterned over several SiC steps, showing PMMA residue particles (white spots, covering about 17% of the surface) and pleats in the graphene (white lines). The scale bar is 2 μm.